\documentstyle[12pt]{article}
\newcommand{\be}{\begin{equation}}
\newcommand{\ee}{\end{equation}}
\newcommand{\ben}{\begin{eqnarray}}
\newcommand{\een}{\end{eqnarray}}   
\begin{document}

\date{MIT-CTP-2789}

\title{Galileo invariant system\\
       and the motion of relativistic {\it d}-branes\thanks{This   
work is supported in part by funds provided by the U.S. Department of   
Energy (D.O.E.) under cooperative research agreement \#DF-FC02-94ER40818,
and by Conselho Nacional de Desenvolvimento Cient\'\i fico e Tecnol\'ogico,
CNPq, Brazil}}

\author{ {D. Bazeia}\\
\\
{\small\it Center for Theoretical Physics, Laboratory for Nuclear Science
and Department of Physics}\\
{\small\it Massachusetts Institute of Technology, Cambridge,
Massachusetts 02139-4307}\\
\\
{\small\it Departamento de F\'\i sica, Universidade Federal da Para\'\i ba}\\
{\small\it Caixa Postal 5008, 58051-970 Jo\~ao Pessoa, Para\'\i ba, Brazil} }

\maketitle

\begin{abstract}
We follow recent work and study the relativistic {\it d}-brane system
in $(d+1,1)$ dimensions and its connection with a Galileo invariant
system in $(d,1)$ dimensions. In particular, we solve {\it d}-brane
systems in $(2,1)$, $(3,1)$ and $(4,1)$ dimensions and show that
their solutions solve the corresponding Galileo invariant systems
in $(1,1)$, $(2,1)$, and $(3,1)$ dimensions. The results are
extended to higher dimensions.
\end{abstract}   
   
\bigskip   
   
PACS number(s): 11.27.+d, 03.40.Gc, 03.65.-w    

\newpage   

\section{Introduction}   

The main objective of this paper is to find solutions of the
relativistic {\it d}-brane system in $(d+1,1)$ dimensions in connection
with solutions of a Galileo invariant system in $(d,1)$ dimensions, as
recently considered in \cite{bja98,jpo98}. Although
we are mainly interested in the cases of $d=1$, $2$ and $3$ spatial
dimensions, some of the results will be naturally extended to $d>1$, arbitrary.
We remark that the Galileo invariant system is defined in $(d,1)$ dimensions,
and is related to the relativistic {\it d}-brane system in one higher spatial
dimension, in $(d+1,1)$ dimensions. The relativistic {\it d}-brane is an
extended object described by $(\phi^0,\phi^1,...,\phi^d)$, where $\phi^0$
is the evolution parameter. The motion of the {\it d}-brane is governed
by the Nambu-Goto action in $(d+1,1)$ space-time dimensions, with
$x^{\mu}=(x^0,{\bf x},x^{d+1})$ and ${\bf x}=(x^1,...,x^d)$. The Galileo
invariant system was recently investigated in the work \cite{bja98,jpo98},
and shows very interesting features, among them direct relation to fluid
mechanics \cite{lli}, to the hydrodynamical formulation of quantum
mechanics \cite{mad26,mer}, and to relativistic membrane \cite{hop82,bho93}
and its generalization to {\it d}-brane in $(d+1,1)$ dimensions.
Other interesting issues are investigated in \cite{hop94,hop95},
in \cite{jev98} and also in \cite{sch96,oga98,sch98}.

The Galileo invariant system is a Lagrangian system governed by the pair of   
fields $\theta=\theta(t,{\bf x})$ and $\rho=\rho(t,{\bf x})$. Its dynamics   
is first-order in time, and the corresponding symplectic structure shows   
that $\theta(t,{\bf x})$ and $\rho(t,{\bf x})$ are canonically   
conjugate \cite{fja88}. There are two equations of motion, one is a
continuity equation for the density $\rho$ and the current
${\bf J}=\rho\,\nabla\theta$ and the other describes first-order time
evolution of $\theta$, and depends on the presence of density-dependent
interactions. The connection between such dynamical system and the membrane
and its generalization to the {\it d}-brane system only appears under the very
specific density-dependent interaction $V(\rho)=\lambda/\rho$.   
   
In the work \cite{bja98} some solutions of the dynamical system were
presented in $(1,1)$ dimensions and in $(d,1)$ for $d>1$. Also, in
\cite{jpo98} the $(1,1)$ dimensional case was shown to be exactly solvable
and solved explicitly. A lesson to be learned here is that this dynamical
system seems to present very specific behavior in $(1,1)$ dimensions,
differently of its general behavior in other dimensions $(d,1)$ for $d>1$.
This fact appears in the former investigations \cite{bja98,jpo98} and also
in the present investigations, which follows an alternate route to find
solutions. We believe that the distinct behavior in $d=1$ is somehow
related to the fact that the current density looses its vectorial character
in one spatial dimension. In this work instead of dealing
directly with the equations of motion that follows from the dynamical system
we solve the {\it d}-brane system in $(d+1,1)$ dimensions. After solving the
{\it d}-brane system we show how the solutions solve the corresponding
Galileo invariant system. Although we shall be concerned mainly with the
{\it d}-brane formulation of the problem, which is of direct interest to
theoretical particle physics, we emphasize that the subject is of broader
interest since it also offers connections to fluid dynamics \cite{lli},
quantum mechanics \cite{mer}, and other
subjects \cite{jev98,sch96,oga98,sch98} of interest to physics.

The present work is organized as follows. In the next
Sec.~{\ref{sec:general}} we introduce the
Galileo invariant system in $(d,1)$ dimensions and the relativistic
general {\it d}-brane system in $(d+1,1)$ dimensions, and we show the interesting
connection between these two apparently distinct systems.
We deal with specific issues in Sec.~{\ref{sec:solutions}}, where we split
the subject into finding solutions of {\it d}-brane systems in $(4,1)$,
$(3,1)$ and in $(2,1)$ dimensions, and showing how they solve the corresponding
Galileo invariant system in $(3,1)$, $(2,1)$, and $(1,1)$ dimensions. In
Sec.~{\ref{sec:generalization}} we generalize results obtained in $d=1,2,3$
to $d>1$, arbitrary. Comments and conclusions are briefly presented in
Sec.~{\ref{sec:comments}}.
   
\section{General considerations}   
\label{sec:general}   
   
Here we introduce the Galileo invariant
system and the relativistic {\it d}-brane system in $(d,1)$ and in $(d+1,1)$
dimensions, respectively. The informations collected below follow in
accordance with Refs.~{\cite{bja98,jpo98}}, and are important for the
issues that compose the rest of the paper. For simplicity we split the
subject into the two subsections that follow.

\subsection{The Galileo invariant system in ({\it d},1) dimensions}

The Galileo invariant system is governed by the action \cite{bja98}
\be
\label{actiong}
I_{\lambda}=\int dt d{\bf x}\left(\theta\,\frac{\partial\rho}{\partial t}-
\frac{1}{2}\,\rho\,\nabla\theta\cdot\nabla\theta-\frac{\lambda}{\rho}\right) 
\ee
Here we have ${\bf x}=(x^1,...,x^d)$, and in this case we are working in
$(d,1)$ dimensions. This system is first-order in time \cite{fja88} and the
fields $\theta(t,{\bf x})$ and $\rho(t,{\bf x})$ are canonical variables,
with the Poisson bracket
\ben
\{\rho(t,{\bf x}),\theta(t,{\bf y})\}=\delta({\bf x}-{\bf y})
\een
The Lagrangian and Hamiltonian are given by
\ben
L&=&\int d{\bf x}\left(\theta\,\frac{\partial\rho}{\partial t}-
\frac{1}{2}\rho\nabla\theta\cdot\nabla\theta-\frac{\lambda}{\rho}\right)
\\
H&=&\int d{\bf x} \left(\frac{1}{2}\rho\nabla\theta\cdot\nabla\theta+
\frac{\lambda}{\rho}\right)
\een
There are two equations of motion. They can be obtained for instance via
\ben
\frac{\partial\theta}{\partial t}&=&\{\theta, H\}
\\
\frac{\partial\rho}{\partial t}&=&\{\rho,H\}
\een
and they have the explicit form
\ben
\label{mg1}
\frac{\partial\theta}{\partial t}+
\frac{1}{2}\,\nabla\theta\cdot\nabla\theta&=&\frac{\lambda}{\rho^2}
\\
\label{mg2}
\frac{\partial\rho}{\partial t}+\nabla\cdot(\rho\,\nabla\theta)&=&0
\een
The first equation of motion couples $\theta$ to $\rho$ via the specific   
density-dependent interaction $V(\rho)=\lambda/\rho$. This equation decouples
$\theta$ from $\rho$ in the absence of interactions, in the limit
$\lambda\to0$. The second equation, Eq.~{(\ref{mg2})}, is a continuity
equation linking the density $\rho$ with the current density
${\bf J}=\rho\nabla\theta$. Here the current density necessarily
couples the two fields $\rho$ and $\theta$. In one spatial
dimension this coupling is non-vectorial, and we believe that this
may be behind the very specific behavior this system presents
in the $(1,1)$ dimensional case.

The hydrodynamical description of quantum mechanics follows with
the Schr\"odinger action
\be
I_S=\int\,dt\,d{\bf x}\left( i\Psi^{*}{\dot\Psi}-
\frac{1}{2}\nabla\Psi^{*}\cdot\nabla\Psi-V_S\right)
\ee
where $V_S=V(t,{\bf x})+{\bar V}(\Psi^{*}\Psi)$ is the potential.
We use
\be
\Psi(t,{\bf x})=\sqrt{\rho}\,e^{i\theta}
\ee
and the specific potential
\be
\label{sp}
V_S=\frac{\lambda}{\rho}-\frac{1}{8}\,\frac{(\nabla\rho)^2}{\rho}
\ee
to make $I_S$ to collapse to $I_{\lambda}$, reproducing the two equations
of motion $(\ref{mg1})$ and $(\ref{mg2})$. The quantum mechanical description
of the system just introduced is then governed by the dynamical system with
the action (\ref{actiong}). This hydrodynamical description is restricted
to pure states \cite{fan57} and one may envisage another route, relying on
the phase-space description of quantum mechanics \cite{wig32}, which seems
appropriate to generalize the above description -- recall that the
(probability) density and the (probability) current density are the
first two velocity moments of the Wigner distribution \cite{cza83}.
On the other hand, we remark that potentials like the one in Eq.~(\ref{sp})
have been considered for instance in Ref.~\cite{dgo94}, and appear under the
assumption of local probability conservation for the usual density and a
current density that is extended to account for diffusion.

Although the dynamical system $(\ref{actiong})$ is
manifestaly invariant under Galileo transformations, it also presents
\cite{bja98} symmetry algebra that can be identified with that of the
Poincar\'e group in $(d+1,1)$ dimensions. To see this, in $(d+1,1)$
dimensions we use light-cone coordinates to identify $(x^0,x^1,...,x^{d+1})$
with $(t,\theta,{\bf x})$, where ${\bf x}$ stands for the transverse
coordinates and
\ben
\label{x+}
t&\equiv&\frac{1}{\sqrt{2}}(x^0+x^{d+1})=x^{+}
\\
\label{x-}
\theta&\equiv&\frac{1}{\sqrt{2}}(x^0-
x^{d+1})=x^{-}
\een
We now introduce the light-cone components of the Poincar\'e   
generators $P^{\mu}$ and $M^{\mu\nu}$ in the usual form   
\ben
P^{\mu}&=&(P^-,P^+,P^i)   
\\
M^{\mu\nu}&=&(M^{+-},M^{+i},M^{-i},M^{ij})   
\een   
To identify the generators of the Poincar\'e symmetry we recall that the
Poincar\'e group contains the extended Galileo group as a
subgroup \cite{sus68}. We write
\ben
\label{p+}
P^-&\equiv&H=\int d{\bf x}\,{\cal E}   
\\
\label{p-}
P^+&\equiv&N=\int d{\bf x} \,\rho   
\\
\label{pi}
P^i&=&\int d{\bf x} \, J^i
\een
and
\ben
\label{+-}
M^{+-}&\equiv&D=\int d{\bf x}\,(t\,{\cal E} - \,\theta\rho)
\\
\label{b+}
M^{+i}&\equiv&B^i=\int d{\bf x}\,\left(t\,J^i- \,\rho\, x^i\right)
\\
\label{b-}
M^{-i}&\equiv&G^i=\int d{\bf x}\,({\cal E} x^i-\,\theta\,J^i)
\\
\label{ang}
M^{ij}&=&\int d{\bf x}\,(x^i\,J^j-x^j\,J^i)
\een
The first set identifies generators of translations, which introduces the   
Hamiltonian $H$, charge $N$ and linear momentum ${\bf P}$. The second set   
identifies generators of rotations, which introduces the dilation $D$,
Galileo boost ${\bf B}$, ${\bf G}$ and the angular momentum ${\bf M}$.
Here we are using
\ben
J^i&=&\rho\,\frac{\partial\theta}{\partial x^i}
\\
{\cal E}&=&\frac{1}{2}\,
\rho\,\nabla\theta\cdot\nabla\theta+\frac{\lambda}{\rho}   
\een
It is not hard to check that the generators $(\ref{p+})$-$(\ref{ang})$ close
the Poincar\'e algebra in $(d+1,1)$ dimensions.

It is interesting to realize that interchangeability of the light-cone   
coordinates $(+)$ and $(-)$ allows interchanging $t$ and $\theta$ in
$(\ref{x+})$ and $(\ref{x-})$, and this further implies interchanging
$P^{+}$ and $P^{-}$ and $M^{+i}$ and $M^{-i}$. We interchange $P^{+}$
and $P^{-}$ with $\rho\leftrightarrow{\cal E}$, which implies
$H\leftrightarrow N$ and ${\bf B}\leftrightarrow-{\bf G}$.   
One verifies that under the interchange of light-cone coordinate the dilation
changes as $D\leftrightarrow -D$, picking up the minus sign in the same   
way the other rotation generators ${\bf B}$ and ${\bf G}$ do. These
properties make the identifications $(\ref{p+})$-$(\ref{p-})$
and $(\ref{+-})$-$(\ref{b-})$ very natural. Also they are important
for constructing the explicit transformations: We offer a simple
illustration by considering the generator of dilation symmetry; here we
know that time changes according to
\be
t\to e^{w}\,t
\ee
To make the interchangeability of light-cone coordinates to work correctly
we must change the $\theta$ field according to
\be
\theta(t,{\bf x})\to e^w\theta(e^w t,{\bf x})
\ee
which is indeed the correct transformation \cite{bja98}.
This reasoning is inspired on unpublished notes \cite{dho98},   
and it helps building all the transformations explicitly, in particular
the apparently misterious field-dependent coordinate transformations
introduced in \cite{bja98}, which now follow naturally from the standard
coordinate and field transformations that appear in a Galileo boost.

There is an alternate way to see the naturalness of the identification of
the generators of the Lorentz symmetry. This was briefly introduced
in \cite{bja98}, and here we make the argument explicit. We introduce the
infinitesimal Lorentz transformations
\be
\delta x^{\mu}=w^{\mu\nu}\,x_{\nu}
\ee
In the $(d+1,1)$ dimensional space-time we change to the former light-cone
coordinates to rewrite the infinitesimal Lorentz transformations in the form
\ben
\delta\,x^{+}&=&w^{+-}\,x^{+}+w^{i+}\,x^{i}
\\
\delta\,x^{-}&=&-\,w^{+-}\,x^{-}+w^{i-}\,x^{i}
\\
\delta\,x^{i}&=&w^{i-}\,x^{+}+w^{i+}\,x^{-}-w^{ij}\,x^{j}
\een
The parameters $w^{+-}$, $w^{i-}$, $w^{i+}$ and $w^{ij}$ respond for dilation,
Galileo boost, field-dependent coordinate transformations and rotations,
respectively. To see this explicitly we notice that the transformations
involving only $w^{+-}$ obey
\ben
\delta\,x^{+}&=&w^{+-}\,x^{+}
\\
\delta\,x^{-}&=&-\,w^{+-}\,x^{-}
\een
We use $w^{+-}=w$ to see that $\delta^{(n)}\,x^{\pm}=(\pm w)^{n}\,x^{\pm}$.
The finite transformations are
\ben
{\tilde x}^{+}&=&e^{w}\,x^{+}
\\
{\tilde x}^{-}&=&e^{-w}\,x^{-}
\een
We use the light-cone identifications $x^+=t$ and
$x^-=\theta_{w}(t,{\bf x})$ and write the new coordinates as ${\tilde x}^{+}=T$
and ${\tilde x}^{-}=\theta(T,{\bf X})$. We then get
\ben
t&\to& T=e^{w}\,t
\\
\theta&\to&\theta_{w}(t,{\bf x})=e^{w}\theta(e^{w}t,{\bf x})
\een
in agreement with the dilation transformations introduced in
Ref.~{\cite{bja98}}.

Let us now focus attention on the transformations introduced by $w^{i-}$,
identifying $w^{i-}=w^{i}$. We obtain
\ben
\delta\,x^{+}&=&0
\\
\delta\,x^{-}&=&w^{i}\,x^{i}
\\
\delta\,x^{i}&=&w^{i}\,x^{+}
\een
We see that
\ben
\delta^{(2)}\,x^{+}&=&0
\\
\delta^{(2)}\,x^{-}&=&\,w^{2}\,x^{+};\hspace{1cm}\delta^{(3)}\,x^{-}=0
\\
\delta^{(2)}\,x^{i}&=&0
\een
The finite transformations are then given by
\ben
T&=&t
\\
X^i&=&x^i+ w^i\,t
\een
and
\be
\theta(T,{\bf X})=\theta_{w}(t,{\bf x})+ w^i x^i+\frac{1}{2}\,w^2\,t
\ee
These transformations identify the Galileo boost -- see Ref.~{\cite{bja98}}.

We now examine the transformations introduced by $w^{i+}$, making the
identification $w^{i+}=w^i$. We get
\ben
\delta\,x^{+}&=&w^{i}\,x^{i}
\\
\delta\,x^{-}&=&0
\\
\delta\,x^{i}&=&w^{i}\,x^{-}
\een
We see that
\ben
\delta^{(2)}\,x^{+}&=&w^{2}\,x^{-};\hspace{1cm}\delta^{(3)}\,x^{+}=0
\\
\delta^{(2)}\,x^{-}&=&0
\\
\delta^{(2)}\,x^{i}&=&0
\een
The finite transformations are given by
\ben
T&=&t+w^i x^i+\frac{1}{2}\,w^2\theta_{w}(t,{\bf x})
\\
X^i&=&x^i+w^i\, \theta_{w}(t,{\bf x})
\een
and also
\be
\theta(T,{\bf X})=\theta_{w}(t,{\bf x})
\ee
In this case we can write
\ben
T&=&t+w^ix^i+\frac{1}{2}\,w^2\theta(T,{\bf X})
\\
X^i&=&x^i+w^i\, \theta(T,{\bf X})
\een
which identify the field-dependent coordinate transformations introduced
in Ref.~{\cite{bja98}}.

The Galileo invariant system also engenders interesting solutions. The
equations of motion $(\ref{mg1})$ and $(\ref{mg2})$ possess
dilation-invariant solutions that can be written in the form
\ben
\label{sthetad}
\theta(t,r)&=&-\frac{r^2}{2(d-1)t}
\\
\label{srhod}
\rho(t,r)&=&\sqrt{\frac{2}{d}\lambda\,}\,(d-1)\,\frac{|t|}{r}
\een
This pair of solutions appears in $(d,1)$ dimensions, for $d>1$, and here
we have set $r=\sqrt{{\bf x}\cdot{\bf x}}$, which identifies the length
of the vector ${\bf x}=(x^1,x^2,...,x^d)$ in $d>1$ spatial dimensions.
In Ref.~{\cite{bja98}} the pair of solutions for $d=2$ and the corresponding
field-dependent coordinate transformations induced by the generators
of ${\bf G}$ were used to obtain other solutions to the dynamical system
in $(2,1)$ dimensions. Here, however, we explore other issues and get the
pair of solutions $(\ref{sthetad})$ and $(\ref{srhod})$ by following an
alternate route, which relies on solving the {\it d}-brane problem in
$(d+1,1)$ dimensions, for $d>1$.

This Galileo invariant system can also be solved in one spatial
dimension, and some solutions were already presented in \cite{bja98}.
Also, in \cite{jpo98} it was shown that the $(1,1)$ dimensional system
is integrable, and the explicit solutions were presented. This result
is confirmed by the investigation that appears in Sec.~{\ref{sec:solutions}}
for the 1-brane system in $(2,1)$ dimensions.

\subsection{The ({\it d}+1,1) dimensional {\it d}-brane system}

We follow Ref.~{\cite{hop82}} to introduce the relativistic
{\it d}-brane system in $(d+1,1)$ dimensions. The {\it d}-brane system is an
extended object described by the coordinates $(\phi^0,\phi^1,...,\phi^d)$,
where $\phi^0$ is the evolution parameter and $(\phi^1,...,\phi^d)$ constitute
the {\it d}-dimensional space that parametrizes the {\it d}-brane. This
object is governed by the Nambu-Goto action
\be
\label{adb}
I_d=-\int d\phi^0 d{\phi^1}...d{\phi^d} \,\, \sqrt{G}
\ee
where $G$ is $(-1)^d$ times the determinant of the induced metric
\be
G_{\alpha\beta}\equiv\frac{\partial x^\mu}{\partial\phi^{\alpha}}\,
\frac{\partial x_\mu}{\partial\phi^{\beta}}
\ee
where $\alpha,\beta=0,1,...,d$. Here the {\it d}-brane is submersed in the
$(d+1,1)$ spacetime and we use light-cone coordinates to represent
$x^{\mu}=(x^0,x^1,...,x^d,x^{(d+1)})$ as $(\tau,\theta,{\bf x})$, where
\ben
\label{tau}
x^{+}&=&\frac{1}{\sqrt{2}}\,(x^0+x^{(d+1)})\equiv\tau
\\
\label{theta}
x^{-}&=&\frac{1}{\sqrt{2}}\,(x^0-x^{(d+1)})\equiv\theta
\een
We use light-cone coordinates with the same motivation of the former case,
where we have shown that the Galileo invariant system in $(d,1)$ dimensions
presents symmetry algebra that can be identified with that of the Poincar\'e
group in $(d+1,1)$ dimensions. This is convenient because light-cone
coordinates introduce the tranverse spatial components
${\bf x}=(x^1,...,x^d)$ very naturally, which we shall identify
directly with the spatial components of the Galileo invariant system 
in $(d,1)$ dimensions.

We identify the evolution parameter $\phi^0$ with the light-cone time $\tau$
$(\tau\equiv\phi^0)$ to write the elements of $G_{\alpha\beta}$ in the form
\ben
G_{00}&=&2\,\frac{\partial\theta}{\partial\tau}-
\frac{\partial{\bf x}}{\partial\tau}\cdot
\frac{\partial{\bf x}}{\partial\tau}
\\
G_{0i}&=&G_{i0}=\frac{\partial\theta}{\partial\phi^i}-
\frac{\partial{\bf x}}{\partial\tau}\cdot
\frac{\partial{\bf x}}{\partial\phi^i}
\\
\label{matrixg}
G_{ij}&=&-g_{ij}=\frac{\partial{\bf x}}{\partial\phi^i}\cdot
\frac{\partial{\bf x}}{\partial\phi^j}
\een
We write
\be
g\equiv\det\left(g_{ij}\right)
\ee
to get
\be
G=g\,\left(2\frac{\partial\theta}{\partial\tau}-
\frac{\partial{\bf x}}{\partial\tau}\cdot
\frac{\partial{\bf x}}{\partial\tau}+ g^{ij}u_i u_j\right)
\ee
where $g^{ik}g_{kj}=\delta^i_j$ and $u_i$ is defined as
\be
\label{defu}
u_i\equiv -\frac{\partial\theta}{\partial\phi^i}+
\frac{\partial{\bf x}}{\partial\tau}\cdot
\frac{\partial{\bf x}}{\partial\phi^i}
\ee
The equations of motion for the {\it d}-brane can be written in the
form \cite{hop82}
\ben
\frac{\partial{\bf x}}{\partial\tau}&=&-\frac{\bf p}{\Pi}+
u^i\frac{\partial{\bf x}}{\partial\phi^i}
\\
\frac{\partial\theta}{\partial\tau}&=&\frac{1}{2\Pi^2}\,
(\,{\bf p}\cdot{\bf p}+g\,)+u^i\frac{\partial\theta}{\partial\phi^i}
\\
\frac{\partial{\bf p}}{\partial\tau}&=&-\frac{\partial}{\partial\phi^i}
\left(\frac{1}{\Pi}\,g\,g^{ij}\frac{\partial{\bf x}}{\partial\phi^j}\right)+
\frac{\partial}{\partial\phi^i}({\bf p}\,u^i)
\\
\frac{\partial\Pi}{\partial\tau}&=&\frac{\partial}{\partial\phi^i}(\Pi\,u^i)
\een
where ${\bf p}$ and $\Pi$ are canonical momenta conjugate to ${\bf x}$
and $\theta$, respectively. The motion is constrained to obey
\be
\label{cg}
{\bf p}\cdot\frac{\partial{\bf x}}{\partial\phi^i}+
\Pi\frac{\partial\theta}{\partial\phi^i}=0
\ee
This condition appears as a consequence of gauge symmetry, which comes
from the freedom to parametrize the {\it d}-brane.

In the light-cone coordinates with the time identification
$\tau\equiv\phi^0$ we can further investigate the physical contents of the
{\it d}-brane after seting $u_i=0$. This condition identifies the
(light-cone) time dependence of the parametrization and allows rewriting
the equations of motion in the simpler form
\ben
{\bf {p}}&=&-\Pi\,\frac{\partial{\bf{x}}}{\partial\tau}
\\
\label{pitau}
\frac{\partial\Pi}{\partial\tau}&=&0
\een
and
\ben
\label{gpi1}
\frac{\partial\theta}{\partial\tau}&=&\frac{1}{2}\,
\left(\frac{\partial{\bf x}}{\partial\tau}\cdot
\frac{\partial{\bf x}}{\partial\tau}+\frac{1}{\Pi^2}\,g\,\right)
\\
\label{gpi2}
\frac{\partial^2{\bf x}}{\partial\tau^2}&=&\frac{1}{\Pi}\,
\frac{\partial}{\partial\phi^i}
\left(\frac{1}{\Pi}\,g\,g^{ij}\frac{\partial{\bf x}}{\partial\phi^j}\right)
\een
Instead of Eq.~(\ref{cg}) the constrained motion now obeys
\be
\label{c}
\frac{\partial\theta}{\partial\phi^i}=\frac{\partial{\bf x}}{\partial\tau}   
\cdot\frac{\partial{\bf x}}{\partial\phi^i}   
\ee
which reproduces Eq.~(\ref{defu}) for $u_i=0$ in a self-consistent way.
The constraint Eq.~(\ref{c}) is independent of the explicit form of the
momentum $\Pi=\Pi({\phi^1,...,\phi^d})$, now $\tau$-independent in
accordance with Eq.~(\ref{pitau}). This fact shows that one is still free
to make $\tau$-independent reparametrization
$(\phi^1,...,\phi^d)\to({\bar\phi}^1,...,{\bar\phi}^d)$ of the {\it d}-brane.
Different choices of the momentum $\Pi$ are related to different choices of
$\tau$-independent parametrization and then we can choose $\Pi$ at
convenience. For instance, the choice $\Pi=-c\,w(\phi^1,...,\phi^d)$
envolving a constant $c$ times a specified function of $\phi^1,...,\phi^d$ is
known \cite{hop82} as the orthonormal gauge. We can illustrate this point by
going from $\Pi({\phi^1,...,\phi^d})$ to $\Pi=-c$, constant, for simplicity.
In this case the above Eqs.~(\ref{gpi1}) and (\ref{gpi2}) become
\ben
\label{g1c}
\frac{\partial^2{\bf x}}{\partial\tau^2}&=&\frac{\partial}{\partial\phi^i}
\left(\frac{1}{c^2}\,{g}\,g^{ij}\frac{\partial{\bf x}}{\partial\phi^j}\right)
\\
\label{g2c}
\frac{\partial\theta}{\partial\tau}&=&\frac{1}{2}
\left(\frac{\partial{\bf x}}{\partial\tau}\cdot
\frac{\partial{\bf x}}{\partial\tau}+ \frac{1}{c^2}\,{g}\,\right)
\een
We now change parametrization by allowing
\be
\frac{\partial{\bf x}}{\partial\phi^i}\to
\frac{\partial{\bf x}}{\partial{\bar\phi}^j}\,
\frac{\partial{\bar\phi}^j}{\partial\phi^i}
\ee
with a similar change in $\theta(\tau,\phi^1,...,\phi^d)$, in accordance
with the constraint Eq.~(\ref{c}). We use
\be
\frac{\partial{\bar\phi}^i}{\partial\phi^j}=c\,\delta^i_j
\ee
as the choice to get rid of the factor $1/c^2$ in
Eqs.~(\ref{g1c}) and (\ref{g2c}). For simplicity we then consider
$\Pi=-1$, which further implies ${\bf p}=\partial{\bf x}/\partial\tau$.
In this case the equations of motion simplify to the equations
\ben
\label{m1}
\frac{\partial^2{\bf x}}{\partial\tau^2}&=&\frac{\partial}{\partial\phi^i} 
\left(g\,g^{ij}\frac{\partial{\bf x}}{\partial\phi^j}\right)
\\
\label{m2}
\frac{\partial\theta}{\partial\tau}&=&\frac{1}{2}
\left(\frac{\partial{\bf x}}{\partial\tau}\cdot
\frac{\partial{\bf x}}{\partial\tau}+ g\,\right)
\een

The constraint Eq.~$(\ref{c})$ was first solved \cite{bho93} in the $d=2$
case, the membrane case and it can also be solved in the general {\it d}-brane
case. We follow Ref.~{\cite{jpo98}} and here the main step concerns
inverting ${\bf x}={\bf x}(\tau,\phi^1,...,\phi^d)$ to get
${\phi^i}={\phi^i}(t,{\bf x})$, after renaming $\tau$ as time $t$.
Here we have
\be
\label{taut}
\frac{\partial}{\partial\tau}=\frac{\partial}{\partial t}+
\nabla\theta\cdot\nabla
\ee
and the constraint Eq.~(\ref{c}) is solved by
\be
\label{csol}
\frac{\partial{\bf x}}{\partial\tau}=\nabla\theta
\ee
In this case Eq.~(\ref{m2}) becomes
\be
\frac{\partial\theta}{\partial t}+
\frac{1}{2}\nabla\theta\cdot\nabla\theta=\frac{1}{2}g
\ee
and reproduces Eq.~(\ref{mg1}) if and only if we further define
\be
\label{defg}
g\equiv\frac{2\lambda}{\rho^2}
\ee

On the other hand the Eq.~(\ref{m1}) can be rewritten as
\be
\label{m1'}
\frac{\partial^2{x^i}}{\partial\tau^2}=\frac{1}{2}
\frac{\partial g}{\partial x^i}
\ee
The proof follows after recognizing that
\be
gg^{ij}=\frac{1}{(d-1)!}\epsilon^{ii_2...i_d}\epsilon^{jj_2...j_d}
\frac{\partial x^{k_2}}{\partial\phi^{i_2}}\cdots
\frac{\partial x^{k_d}}{\partial\phi^{i_d}}
\frac{\partial x^{k_2}}{\partial\phi^{j_2}}\cdots
\frac{\partial x^{k_d}}{\partial\phi^{j_d}}
\ee
In this case we can use
\be
\epsilon^{jj_2...j_d}\frac{\partial x^{k_2}}{\partial\phi^{j_2}}\cdots
\frac{\partial x^{k_d}}{\partial\phi^{j_d}}
\frac{\partial x^{k}}{\partial\phi^{j}}
\equiv\epsilon^{kk_2...k_d}\sqrt{g}
\ee
to get to
\be
gg^{ij}\frac{\partial x^{k}}{\partial\phi^{j}}=\frac{1}{(d-1)!}
\epsilon^{ii_2...i_d}
\epsilon^{kk_2...k_d}\frac{\partial x^{k_2}}{\partial\phi^{i_2}}\cdots
\frac{\partial x^{k_d}}{\partial\phi^{i_d}}\sqrt{g}
\ee
and now we can write, recalling the smoothness of
${\bf x}={\bf x}(\tau,\phi^1,...,\phi^d)$ and (anti) symmetry of the
Levi-Civita symbol,
\ben
\frac{\partial}{\partial\phi^i}\left(gg^{ij}
\frac{\partial x^{k}}{\partial\phi^{j}}\right)&=&
\frac{1}{(d-1)!}\epsilon^{ii_2...i_d}\epsilon^{kk_2...k_d}
\frac{\partial x^{k_2}}{\partial\phi^{i_2}}\cdots
\frac{\partial x^{k_d}}{\partial\phi^{i_d}}
\Biggl[\frac{\partial}{\partial\phi^{i}}\left(\sqrt{g}\right)\Biggr]
\nonumber\\
&=&\frac{1}{(d-1)!}\epsilon^{ii_2...i_d}\epsilon^{kk_2...k_d}
\frac{\partial x^{k_2}}{\partial\phi^{i_2}}\cdots
\frac{\partial x^{k_d}}{\partial\phi^{i_d}}\Biggl[
\frac{\partial}{\partial x^{l}}\left(\sqrt{g}\right)\Biggr]
\frac{\partial x^{l}}{\partial\phi^{i}}\nonumber\\
&=&\frac{1}{(d-1)!}\epsilon^{lk_2...k_d}\epsilon^{kk_2...k_d}
\sqrt{g}\Biggl[\frac{\partial}{\partial x^{l}}
\left(\sqrt{g}\right)\Biggr]
\een
We use the identity
\be
\frac{1}{(d-1)!}\epsilon^{ik_2...k_d}
\epsilon^{jk_2...k_d}\equiv{\delta^{ij}}
\ee
to obtain
\be
\frac{\partial}{\partial\phi^i}\left(gg^{ij}
\frac{\partial x^{k}}{\partial\phi^{j}}\right)
= \frac{1}{2}\frac{\partial g}{\partial x^{k}}
\ee
which ends the proof. We then use Eqs.~(\ref{taut}) and (\ref{csol}) to write
\ben
\frac{\partial^2{\bf x}}{\partial\tau^2}&=&
\frac{\partial}{\partial\tau}\nabla\theta\nonumber
\\
&=&\nabla\frac{\partial\theta}{\partial t}+
\nabla\theta\cdot\nabla(\nabla\theta)\nonumber
\\
&=&\nabla\left(\frac{\partial\theta}{\partial t}+
\frac{1}{2}\nabla\theta\cdot\nabla\theta\right)
\een
We use this together with Eq.~(\ref{m1'}) to obtain, discarding an
unimportant constant,
\be
\frac{\partial\theta}{\partial t}+
\frac{1}{2}\nabla\theta\cdot\nabla\theta=\frac{1}{2}g
\ee
This equation also reproduces the Eq.~(\ref{mg1}) for $g$ defined by
Eq.~(\ref{defg}).

The Galileo invariant system $(\ref{actiong})$ contains another equation of
motion. It is Eq.~(\ref{mg2}), the continuity equation. It is
obtained from the relativistic {\it d}-brane system as follows. We have,
by definition
\be
\frac{\partial g}{\partial\tau}\equiv
g\,g^{ij}\frac{\partial}{\partial\tau}g_{ij}
\ee
Also
\be
\frac{\partial g}{\partial\tau}=\frac{\partial g}{\partial t}+
\nabla\theta\cdot\nabla g
\ee
We use these results together with
\ben
g^{ij}\frac{\partial}{\partial\tau}g_{ij}&=&
\frac{\partial\phi^i}{\partial x^k}\frac{\partial\phi^j}{\partial x^k}
\frac{\partial}{\partial\tau}\left(\frac{\partial x^l}{\partial\phi^i}
\frac{\partial x^l}{\partial\phi^j}\right)\nonumber\\
&=&2\frac{\partial\phi^j}{\partial x^k}\frac{\partial}{\partial\phi^j}
\frac{\partial x^k}{\partial\tau}\nonumber\\
&=&2\frac{\partial}{\partial x^k}\frac{\partial\theta}{\partial x^k}
\een
to obtain
\be
\frac{\partial g}{\partial t}+\nabla\theta\cdot\nabla g=2g\nabla^2\theta
\ee
which reproduces the continuity equation after choosing $g=2\lambda/\rho^2$,
as given by Eq.~{(\ref{defg})}. The motion of the {\it d}-brane in $(d+1,1)$   
dimensions is then governed by the equations of motion that describe the
dynamical system $(\ref{actiong})$ in $(d,1)$ dimensions. We remark
that the above proof only works under the definition introduced in
Eq.~{(\ref{defg})}, with the density-dependent interaction described
by the very specific potential $V(\rho)=\lambda/\rho$.

There is an alternate way to make the {\it d}-brane system in $(d+1,1)$
dimensions to reproduce the Galileo invariant system in $(d,1)$ dimensions.
It follows like in Ref.~{\cite{hop94}}, for instance, but now in $d+1$
spatial dimensions. It relies on reformulating the description of the
{\it d}-brane according to the identifications \cite{hpri}
\ben
x^{+}&=&\frac{1}{\sqrt{2}}\,(x^0+x^{d+1})\equiv t=\phi^0
\\
x^{-}&=&\frac{1}{\sqrt{2}}\,(x^0-x^{d+1})\equiv\theta
\een
and also
\be
\phi^i\equiv x^i~,\hspace{1cm}i=1,2,...,d
\ee
This is interesting since now $\theta=\theta(t,{\bf x})$ is directly identified
with the $\theta$ field of the Galileo invariant system in $(d,1)$ dimensions.
To see this explictly we recall from Eq.~(\ref{adb}) that the Lagrangian
density can be written as
\be
{\cal L}_d=\sqrt{G}
\ee
and now
\be
G=2\,\frac{\partial\theta}{\partial t}+\nabla\theta\cdot\nabla\theta
\ee
The momentum conjugate to $\theta$ is given by
\be
\Pi=\frac{1}{\sqrt{2\,\frac{\partial\theta}{\partial t}+
\nabla\theta\cdot\nabla\theta}}
\ee
and this gives the Hamiltonian density
\be
{\cal H}=-\frac{1}{2\Pi}-\frac{1}{2}\,\Pi\,\nabla\theta\cdot\nabla\theta
\ee
We use this Hamiltonian density to write the first-order Lagrangian density,
after defining $\Pi\equiv-\rho/\sqrt{2\lambda}$. We discard a total
time derivative and ignore an unimportant multiplicative constant to obtain
\be
{\cal L}=\theta\,\frac{\partial\rho}{\partial t}-
\frac{1}{2}\,\rho\,\nabla\theta\cdot\nabla\theta-\frac{\lambda}{\rho}
\ee
This is exactly the Lagrangian density that follows from the action
$(\ref{actiong})$ that defines the Galileo invariant system in $(d,1)$
dimensions.

\section{Some brane solutions}
\label{sec:solutions}

Let us now deal with solutions of {\it d}-brane systems in
$(2,1)$, $(3,1)$ and $(4,1)$ dimensions, and with the relations between such
solutions and solutions of the Galileo invariant system in one, two and
three spatial dimensions. We split the subject into the three
subsections that follow, which deal with $d=1,2,3$ separately.

\subsection{Solutions for {\it d}=1}   

Here we consider the relativistic 1-brane system in $(2,1)$ dimensions.
We parametrize the system with $\phi$, and the unidimensional transverse
coordinate is given by $x=x(\tau,\phi)$. Also, the matrix $g_{ij}=
(\partial{\bf x}/\partial\phi^i)\cdot(\partial{\bf x}/\partial\phi^j)$
and its determinant $g$ now become the very same thing, explicitly 
\be
\label{g}
g=\left(\frac{\partial x}{\partial\phi}\right)^2
\ee
Also, $g\,g^{ij}\to1$ and the equation of motion $(\ref{m1})$ becomes
\be
\label{wave}
\frac{\partial^2 x}{\partial\tau^2}-\frac{\partial^2 x}{\partial \phi^2}=0   
\ee
The other Eqs.~(\ref{c}) and (\ref{m2}) give
\begin{eqnarray}
\label{thetat}
\frac{\partial\theta}{\partial\phi}&=&
\frac{\partial x}{\partial\tau}\,\frac{\partial x}{\partial\phi}
\\
\label{thetaphi}
\frac{\partial\theta}{\partial\tau}&=&
\frac{1}{2}\Biggl[\left(\frac{\partial x}{\partial\tau}\right)^2+
\left(\frac{\partial x}{\partial\phi}\right)^2\Biggr]
\end{eqnarray}

We can solve Eq.~$(\ref{wave})$ directly. It is a wave equation and presents
the general solution
\be
\label{sol}
x(\tau,\phi)=f_{+}(\tau+\phi)+f_{-}(\tau-\phi)
\ee
The solution is written in terms of two arbitrary functions $f_{\pm}$, and
here we have
\be
\label{xtau}
\frac{\partial x}{\partial\tau}=f'_{+}(\tau+\phi) + f'_{-}(\tau-\phi)
\ee
and
\be
\label{xphi}
\frac{\partial x}{\partial\phi}=f'_{+}(\tau+\phi) - f'_{-}(\tau-\phi)
\ee
where we are using the notation $f'(z)=(\partial f/\partial z)$. We use these
expressions to rewrite Eqs.~$(\ref{thetat})$ and $(\ref{thetaphi})$ as
\ben
\frac{\partial\theta}{\partial\tau}&=&[f'_{+}(\tau+\phi)]^2+
[f'_{-}(\tau-\phi)]^2
\\
\label{thetaphi1}
\frac{\partial\theta}{\partial\phi}&=&[f'_{+}(\tau+\phi)]^2-
[f'_{-}(\tau-\phi)]^2
\een
in order to get
\begin{eqnarray}
\frac{1}{2}\left(\frac{\partial}{\partial\tau}+
\frac{\partial}{\partial\phi}\right)\theta(\tau,\phi) &=&
[f'_{+}(\tau+\phi)]^2 
\\
\frac{1}{2}\left(\frac{\partial}{\partial\tau}-   
\frac{\partial}{\partial\phi}\right)\theta(\tau,\phi) &=&
[f'_{-}(\tau-\phi)]^2
\end{eqnarray}
These results allow writing $\theta(\tau,\phi)=\theta_+(\tau+\phi)+   
\theta_-(\tau-\phi)$, and the general solution for $\theta$ can be written
in the form
\be
\label{thetatauphi1}
\theta(\tau,\phi)=\int^{(\tau+\phi)}\,[f'_{+}(z)]^2\,dz+
\int^{(\tau-\phi)}\, [f'_{-}(z)]^2\,dz
\ee
The results given by Eqs.~(\ref{sol}) and (\ref{thetatauphi1}) constitute the
pair of general solutions ${\bf x}(\tau,\phi)$ and
$\theta(\tau,\phi)$ of the 1-brane system.

To connect the 1-brane system to the Galileo invariant system in $(1,1)$
dimensions we follow the general investigations introduced in
Sec.~{\ref{sec:general}}. We start by rewriting $\theta(\tau,\phi)$ as
$\theta(t,x)$. Since we already have $x=x(\tau,\phi)$ in Eq.~$(\ref{sol})$,
we write $\phi=h(\tau,x)$ and change $\tau\to t$ to get to
\be
x\equiv f_{+}(t+h(t,x))+f_{-}(t-h(t,x))
\ee
in a way such that
\be
(1+h_t)\,f'_{+}+(1-h_t)\,f'_{-}=0
\ee
and
\be
h_x\, f'_{+}-h_x \,f'_{-}=1
\ee
We also get
\be
h_t=\frac{\partial h}{\partial t}=-\,\frac{f'_{+}+f'_{-}}{f'_{+}-f'_{-}}
\ee
and
\be
h_x=\frac{\partial h}{\partial x}=\frac{1}{f'_{+}-f'_{-}}
\ee
These results allow writing
\be
\label{thetax}
\theta(t,x)=\int^{[t+h(x,t)]}\,[f'_{+}(z)]^2\,dz\,+\,
\int^{[t-h(x,t)]}\, [f'_{-}(z)]^2\,dz
\ee
On the other hand, we use Eq.~{$(\ref{defg})$} to get
\be
g=\left(\frac{\partial x}{\partial\phi}\right)^2=\frac{2\lambda}{\rho^2}
\ee
and this leads to the result
\be
\label{rho}
\rho(t,x)=\pm\,\frac{\sqrt{2\lambda}}{f'_{+}-f'_{-}}
\ee
The above field configurations $\theta(t,x)$ and $\rho(t,x)$ obey
the pair of equations
\ben
\label{motion1}
\frac{\partial\rho}{\partial t}+
\frac{\partial}{\partial x}\Biggl[\rho\left(\frac{\partial\theta}   
{\partial x}\right)\Biggr]&=&0
\\
\label{motion2}   
\frac{\partial\theta}{\partial t}+   
\frac{1}{2}\left(\frac{\partial\theta}{\partial x}\right)^2&=&
\frac{\lambda}{\rho^2}
\een
which are the equations of motion $(\ref{mg1})$ and $(\ref{mg2})$ in the   
$(1,1)$ dimensional case.

Furthermore, we can use Eq.~$(\ref{thetax})$ to write
\be
\label{thetat1}
\frac{\partial\theta}{\partial t}=-2\,f'_+\,f'_-
\ee
Also, the current density $J(x,t)=\rho\,(\partial\theta/\partial x)$ can be
written as
\be
\label{current}
J(t,x)=\pm\,\sqrt{2\lambda}\,\,\frac{f'_+ + f'_-}{f'_+ -f'_-}
\ee
It is now interesting to see that if we set $f_+$ or $f_-$ to zero,
that is, if we solve Eq.~$(\ref{wave})$ with only one of the two independent
solutions we get to the result that both $\theta$ and $\rho$ are
time-independent: We see that $\theta$ is time-independent directly from
Eq.~$(\ref{thetat1})$; from Eq.~$(\ref{current})$ we get that the current
density is constant, and so the continuity equation demands the density to
be time-independent. We remark that the equations of motion $(\ref{motion1})$
and $(\ref{motion2})$ impose that the current density is a specific constant
\be
J(t,x)=\pm\,\sqrt{2\lambda}
\ee
when one considers time-independent configurations. This means that the
particular solution $x(\tau,\phi)=f_+(\tau+\phi)$ or
$x(\tau,\phi)=f_-(\tau-\phi)$ to the Eq.~$(\ref{wave})$ reproduces all the
static solutions of the dynamical system governed by the action
$(\ref{actiong})$ in $(1,1)$ dimensions

For the $(1,1)$ dimensional system the energy of static configurations
$\theta=\theta(x)$ and $\rho=\rho(x)$ can be written in the
form
\begin{eqnarray}   
E&=&\int dx \Biggl[\,\frac{1}{2}\,\rho\left(\frac{d\theta}{dx}
\right)^2+\frac{\lambda}{\rho}\,\Biggr]\nonumber\\   
&=&\pm\,\sqrt{2\lambda}\int dx \,\frac{d\theta}{dx}+   
\frac{1}{2}\int dx\, \rho\left(\frac{d\theta}{dx}\mp   
\frac{\sqrt{2\lambda}}{\rho}\,\right)^2   
\end{eqnarray}   
and so is minimized to the value   
\be   
E_M=\pm\,\sqrt{2\lambda} \,\,[\theta(x=\infty)-\theta(x=-\infty)]   
\ee   
for field configurations that obey   
\be   
\frac{d\theta}{dx}=\pm   
\frac{\sqrt{2\lambda}}{\rho}   
\ee   
which are the same solutions with constant and uniform current density   
already obtained. Here we notice that the energy can also be written as   
\be   
E_M=\int dx\, \rho\,\left(\frac{d\theta}{dx}\right)^2=   
2\int dx\,\, \frac{\lambda}{\rho}   
\ee   
This shows that the kinetic and potential parts of the energy contritube   
equally, as usually happens with BPS solutions \cite{bps}. This means that
the system described by the Eqs.~$(\ref{motion1})$ and $(\ref{motion2})$
is the bosonic portion of some supersymmetric system, and that the 1-brane
solutions with $(\tau+\phi)$ or $(\tau-\phi)$ are the solutions that generate
all the BPS solutions of the corresponding dynamical system described by the
action $(\ref{actiong})$ in $(1,1)$ dimensions.

\subsection{Solutions for {\it d}=2}

Let us consider the $(3,1)$ dimensional problem. This is the case with
$d=2$, the membrane case where $(\phi^1,\phi^2)=(\phi,\psi)$ and
${\bf x}=(x,y)$. This problem was already considered in
\cite{hop82,bho93,hop94}. Here we have
\ben
g_{11}&=&\left(\frac{\partial x}{\partial\phi}\right)^2+
\left(\frac{\partial y}{\partial\phi}\right)^2
\\
g_{12}&=&g_{21}=\frac{\partial x}{\partial\phi}\frac{\partial x}{\partial\psi}+
\frac{\partial y}{\partial\phi}\frac{\partial y}{\partial\psi}   
\\
g_{22}&=&\left(\frac{\partial x}{\partial\psi}\right)^2+
\left(\frac{\partial y}{\partial\psi}\right)^2   
\een
and we get
\be
g=\left(\frac{\partial x}{\partial\phi}\frac{\partial y}{\partial\psi}-   
\frac{\partial x}{\partial\psi}\frac{\partial y}{\partial\phi}\right)^2   
\ee
We can write $g=\{x,y\}^2$, where  
\be   
\{x,y\}=\frac{\partial x}{\partial\phi}\frac{\partial y}{\partial\psi}-   
\frac{\partial x}{\partial\psi}\frac{\partial y}{\partial\phi}   
\ee
is the Poisson bracket with respect to the membrane coordinates $(\phi,\psi)$. 
We remark that such identification is only possible in the membrane case,
for $d=2$.

In Ref.~{\cite{bja98}} one finds solutions of the Galileo invariant system in
the planar ($d=2$) case, which present dilation and circular symmetries. For
this reason we choose the circular {\it Ansatz}    
\ben
\label{x2}
x&=&R(\tau,\phi)\,\cos(\psi)
\\
\label{y2}
y&=&R(\tau,\phi)\,\sin(\psi)   
\een
In this case we get
\ben
g_{11}&=&\left(\frac{\partial R}{\partial\phi}\right)^2   
\\
g_{12}&=&g_{21}=0   
\\
g_{22}&=&R^2   
\een
which gives
\be
\label{g2}
g=R^2\left(\frac{\partial R}{\partial\phi}\right)^2   
\ee
We also have
\ben
g\,g^{11}&=&R^2   
\\
g\,g^{12}&=&g\,g^{21}=0   
\\
g\,g^{22}&=&\left(\frac{\partial R}{\partial\phi}\right)^2   
\een

We use the equation of motion $(\ref{m1})$ to get from the equations for $x$
and $y$ the same equation for $R=R(\tau,\phi)$
\be
\label{R2}
\frac{\partial^2 R}{\partial \tau^2}=R^2\,\frac{\partial^2R}{\partial\phi^2}+
R\,\left(\frac{\partial R}{\partial\phi}\right)^2
\ee
This equation is solved by separating variables. Here we get
\be
\label{R2s}
R(\tau,\phi)=2^{1/2}\,\frac{\phi}{\tau}   
\ee   
The other Eqs.~$(\ref{c})$ and $(\ref{m2})$ give   
\be
\label{theta2s}   
\theta(\tau,\phi)=-\frac{\phi^2}{\tau^3}   
\ee
We use Eqs.~(\ref{x2}) and (\ref{y2}) to write
\ben
\frac{\partial x}{\partial\tau}\frac{\partial x}{\partial\phi}+
\frac{\partial y}{\partial\tau}\frac{\partial y}{\partial\phi}&=&
\left(\frac{\partial R}{\partial\tau}\right)
\left(\frac{\partial R}{\partial\phi}\right)
\\
\frac{\partial x}{\partial\tau}\frac{\partial x}{\partial\psi}+
\frac{\partial y}{\partial\tau}\frac{\partial y}{\partial\psi}&=&0
\een
Also, from Eq.~(\ref{R2s}) we get
\ben
\frac{\partial R}{\partial\tau}&=&-2^{1/2}\,\frac{\phi}{\tau^2}
\\
\frac{\partial R}{\partial\phi}&=&2^{1/2}\,\frac{1}{\tau}
\een
and now Eq.~(\ref{theta2s}) allows checking that
\ben
\frac{\partial\theta}{\partial\phi}&=&-2\,\frac{\phi}{\tau^3}=
\left(\frac{\partial R}{\partial\tau}\right)
\left(\frac{\partial R}{\partial\phi}\right)
\\
\frac{\partial\theta}{\partial\psi}&=&0
\een
in explicit agreement with the constraint Eq.~(\ref{c}), as expected.

We notice that $R^2=x^2+y^2$ and this allows getting   
\be   
\phi^2=\frac{1}{2}\,\tau^2\,(x^2+y^2)   
\ee   
and so we can write   
\be
\theta(t,x,y)=-\frac{1}{2 t}(x^2+y^2)   
\ee
We also have
\be
g=\frac{2}{\tau^2}\,R^2   
\ee
and now Eq.~$(\ref{defg})$ allows getting   
\be
\rho(t,x,y)=\frac{\sqrt{\lambda}\,\,|t|}{\sqrt{(x^2+y^2)}}
\ee
This pair of solutions identifies the circularly symmetric, dilation
invariant solutions presented in \cite{bja98} for the Galileo invariant
system in $(2,1)$ dimensions.

We remark that the above solution $(\ref{R2s})$ is a special
case of the solution found in \cite{hop94}, which gives another pair of
solution to the Galileo invariant system, a pair that presents no dilation
invariance anymore. This fact was already known to the author of
Ref.~{\cite{hop94}}.

\subsection{Solutions for {\it d}=3}

Let us now consider the 3-brane system in $(4,1)$ dimensions. We
parametrize this 3-brane with $(\phi^1,\phi^2,\phi^3)=(\phi,\chi,\psi)$ in
space ${\bf x}=(x,y,z)$. We choose the spherical {\it Ansatz}
\ben
x&=&R(\tau,\phi)\,\sqrt{1-\chi^2}\,\cos(\psi)   
\\
y&=&R(\tau,\phi)\,\sqrt{1-\chi^2}\,\sin(\psi)   
\\
z&=&R(\tau,\phi)\,\chi   
\een
In this case we use Eq.~{(\ref{matrixg})} to obtain   
\ben
g_{11}&=&\left(\frac{\partial R}{\partial\phi}\right)^2
\\
g_{22}&=&\frac{R^2}{1-\chi^2}   
\\
g_{33}&=&R^2\,\left(1-\chi^2\right)   
\een   
with $g_{ij}=0$ for $i\neq j$. This implies that   
\be
\label{g3}
g=R^4\,\left(\frac{\partial R}{\partial\phi}\right)^2   
\ee
We then get $g\,g^{ij}=0$ for $i\neq j$, and for the diagonal elements we
have   
\ben   
g\,g^{11}&=&R^4   
\\
g\,g^{22}&=&(1-\chi^2)\,R^2\,\left(\frac{\partial R}{\partial\phi}\right)^2   
\\   
g\,g^{33}&=&\frac{R^2}{1-\chi^2}
\left(\frac{\partial R}{\partial\phi}\right)^2   
\een   
We use Eq.~{(\ref{m1})} to obtain from the equations for $x$, $y$, and $z$
the same equation for $R=R(\tau,\phi)$   
\be
\label{R3}   
\frac{\partial^2 R}{\partial\tau^2}=R^4\frac{\partial^2 R}{\partial\phi^2}+   
2\,R^3\,\left(\frac{\partial R}{\partial\phi}\right)^2   
\ee

This equation for $R(\tau,\phi)$ can also be solved by separating variables.
We see that   
\be
\label{R3d}
R(\tau,\phi)=3^{1/4}\left(\frac{\phi}{\tau}\right)^{1/2}
\ee
explicitly solves Eq.~{(\ref{R3})}. This result can be used in
Eqs.~(\ref{c}) and (\ref{m2}) to give
\be
\label{theta3s}
\theta(\tau,\phi)=-\frac{1}{4}\,3^{1/2}\,\frac{\phi}{\tau^2}
\ee
We use the spherical {\it Ansatz} to write
\ben
\frac{\partial x}{\partial\tau}\frac{\partial x}{\partial\phi}+
\frac{\partial y}{\partial\tau}\frac{\partial y}{\partial\phi}+
\frac{\partial z}{\partial\tau}\frac{\partial z}{\partial\phi}&=&
\left(\frac{\partial R}{\partial\tau}\right)
\left(\frac{\partial R}{\partial\phi}\right)
\\
\frac{\partial x}{\partial\tau}\frac{\partial x}{\partial\psi}+
\frac{\partial y}{\partial\tau}\frac{\partial y}{\partial\psi}+
\frac{\partial z}{\partial\tau}\frac{\partial z}{\partial\psi}&=&0
\\
\frac{\partial x}{\partial\tau}\frac{\partial x}{\partial\chi}+
\frac{\partial y}{\partial\tau}\frac{\partial y}{\partial\chi}+
\frac{\partial z}{\partial\tau}\frac{\partial z}{\partial\chi}&=&0
\een
Also, from Eq.~(\ref{R3d}) we get
\ben
\frac{\partial R}{\partial\tau}&=&-\frac{1}{2}\,3^{1/4}
\left(\frac{\tau}{\phi}\right)^{1/2}\,\frac{\phi}{\tau^2}
\\
\frac{\partial R}{\partial\phi}&=&\frac{1}{2}\,3^{1/4}
\left(\frac{\tau}{\phi}\right)^{1/2}\,\frac{1}{\tau}
\een
We use Eq.~(\ref{theta3s}) to see that
\be
\frac{\partial\theta}{\partial\phi}=-\frac{1}{4}\,3^{1/2}\,\frac{1}{\tau^2}
\ee
which shows explicit agreement with the constraint Eq.~(\ref{c}).

To obtain $\theta$ in terms of the variables $(t,x,y,z)$ we notice that
the spherical {\it Ansatz} allows writing
\be
x^2+y^2+z^2=R^2=3^{1/2}\,\frac{\phi}{\tau}
\ee
We use this result in Eq.~{(\ref{theta3s})} to obtain
\be
\label{theta3}
\theta(t,x,y,z)=-\frac{1}{4t}\left(x^2+y^2+z^2\right)
\ee
The density is obtained with Eqs.~(\ref{defg}) and (\ref{g3}). It reads
\be
\label{rho3}
\rho(t,x,y,z)=2\sqrt{\frac{2}{3}\,\lambda}\frac{|t|}{\sqrt{x^2+y^2+z^2}}
\ee
and together with Eq.~(\ref{theta3}) forms a pair of solutions of the
Galileo invariant system $(\ref{actiong})$ in (3,1) dimensions.
This pair of solutions is exactly the dilation invariant solutions
found in \cite{bja98} for the Galileo invariant system in $d=3$,
with the density-dependent interaction $V(\rho)=\lambda/\rho$. They are the
solutions $(\ref{sthetad})$ and $(\ref{srhod})$ in the case $d=3$.
 
\section{Generalization}
\label{sec:generalization}

The results obtained in Sec.~{\ref{sec:solutions}} for the {\it d}-brane
system in $d=2$ and $3$ can be naturally extended to dimensions higher than
$d=3$. However, instead of generalizing the former spherical {\it Ansatz}
to arbitrary dimension we follow another route, which concerns generalizing
the results we have already obtained in the former Sec.~{\ref{sec:solutions}}.
We do this by first writing Eqs.~$(\ref{R2})$ and $(\ref{R3})$ together   
\ben
\frac{\partial^2 R}{\partial \tau^2}&=&
R^2\,\frac{\partial^2R}{\partial\phi^2}+
R\,\left(\frac{\partial R}{\partial\phi}\right)^2\hspace{1cm}(d=2)
\\
\frac{\partial^2 R}{\partial\tau^2}&=&R^4\frac{\partial^2 R}{\partial\phi^2}+
2\,R^3\,\left(\frac{\partial R}{\partial\phi}\right)^2\hspace{1cm}(d=3)  
\een
These results suggest the general behavior   
\be
\label{Rg}
\frac{\partial^2 R}{\partial\tau^2}=
R^{2(d-1)}\frac{\partial^2R}{\partial\phi^2}+   
(d-1)\,R^{2(d-1)-1}\left(\frac{\partial R}{\partial\phi}\right)^2   
\ee
This equation is nonlinear in all but the $d=1$ case,
where it reproduces the wave equation already considered in
the former Sec.~{\ref{sec:solutions}}.

Fortunately, we can solve Eq.~(\ref{Rg}) explicitly in the general $d>1$
case. We have the solution
\be
\label{Rds}
R(\tau,\phi)=\left(d^{1/2}\,\frac{\phi}{\tau}\right)^{1/(d-1)}
\ee
On the other hand, we also have the results
\ben
g&=&R^2\,\left(\frac{\partial R}{\partial\phi}\right)^2\hspace{1cm}(d=2)
\\
g&=&R^4\,\left(\frac{\partial R}{\partial\phi}\right)^2\hspace{1cm}(d=3)
\een
They suggest the general behavior
\be
\label{gg}
g=R^{2(d-1)}\,\left(\frac{\partial R}{\partial\phi}\right)^2
\ee
The results for $\theta$ for $d=2$ and $3$ are given by Eqs.~(\ref{theta2s})
and (\ref{theta3s}). They can be rewritten as
\ben
\theta(\tau,\phi)&=&-\,\frac{\phi^2}{\tau^3}=
-\frac{1}{2}\,\frac{1}{\tau}\,\left(2^{1/2}\,\frac{\phi}{\tau}\right)^2
\hspace{1cm}(d=2)
\\
\theta(\tau,\phi)&=&-\frac{1}{4}\,3^{1/2}\,\frac{\phi}{\tau^2}=
-\frac{1}{2}\,\frac{1}{2\tau}\,\Biggl[3^{1/4}
\left(\frac{\phi}{\tau}\right)^{1/2}\,\Biggr]^2 \hspace{1cm}(d=3)
\een
They and the result given by Eq.~{(\ref{Rds})} for $R(\tau,\phi)$
allow the following generalization
\be
\theta(\tau,\phi)=-\frac{1}{2}\,\frac{R^2}{(d-1)\tau}
\ee
which is valid under the restriction $d>1$.
For the density we use Eqs.~(\ref{defg}) and (\ref{gg}) to get
\be
\rho(\tau,\phi)=\sqrt{\frac{2\lambda}{d}}(d-1)\frac{|\tau|}{R}
\ee
These results can be rewritten as
\ben
\theta(t,{\bf x})=-\frac{1}{2}\,\frac{r^2}{(d-1)t}
\\
\rho(t,{\bf x})=\sqrt{\frac{2\lambda}{d}\,}(d-1)\frac{|t|}{r}
\een
after changing $(\tau,\phi)\to (t,{\bf x})$ and using
$r=\sqrt{{\bf x}\cdot{\bf x}}$. They are the results $(\ref{sthetad})$
and $(\ref{srhod})$, found in Ref.~{\cite{bja98}} for the Galileo invariant
system $(\ref{actiong})$ in $(d,1)$ dimensions, for $d>1$.

There are at least two other interesting issues related to Eq.~(\ref{Rg}).
The first one is that it can be obtained from the Lagrangian density
\be
{\cal L}=\frac{1}{2}\left(\frac{\partial R}{\partial\tau}\right)^2-
\frac{1}{2}\,R^{2(d-1)}\,\left(\frac{\partial R}{\partial\phi}\right)^2
\ee
which can be rewritten in the form
\be
{\cal L}=\frac{1}{2}\,{\cal G}^{\alpha\beta}\partial_{\alpha}R\,
\partial_{\beta}R
\ee
where we are using that
\be
\partial_{\alpha}=\left(\frac{\partial}{\partial\tau},
\frac{\partial}{\partial\phi}\right)
\ee
and
\ben
{\cal G}^{00}&=&1
\\
{\cal G}^{01}&=&{\cal G}^{10}=0
\\
{\cal G}^{11}&=&-R^{2(d-1)}
\een
in the effective $(1,1)$ dimensional space-time $(\tau,\phi)$. Here we
introduce the determinant ${\cal G}=\det({\cal G}^{\alpha\beta})=-R^{2(d-1)}$.
In general it depends on the point $(\tau,\phi)$, but in the $d=1$ case
it becomes constant, independent of both the $\phi$ and $\tau$ coordinates
of the bidimensional space-time.

The second issue related to Eq.~(\ref{Rg}) concerns the case of
considering $\tau$-independent spherical {\it Ansatz}, that is, of reducing
$R(\tau,\phi)$ to $R(\phi)$. In this case Eq.~(\ref{Rg}) becomes
\be
\label{Rgphi}
R\,\frac{d^2R}{d\phi^2}+(d-1)\,
\left(\frac{dR}{d\phi}\right)^2=0
\ee
A direct consequence of $R$ being $\tau$-independent is that
\be
\frac{\partial g}{\partial\tau}=0
\ee
Furthermore, from Eqs.~(\ref{gg}) and (\ref{Rgphi}) we obtain
\be
\frac{\partial g}{\partial\phi}=0
\ee
and so $g$ is constant. From Eq.~(\ref{m2}) we then get
\be
\label{thetatind}
\theta=\frac{1}{2}\,g\,\tau
\ee
which is independent of $\phi$ and so compatible with the constraint
Eq.~(\ref{c}). An explicit  illustration is given after solving
Eq.~(\ref{Rgphi}). Here we see that
\be
\label{Rgenind}
R(\phi)=\phi^{1/d}
\ee
solves Eq.~(\ref{Rgphi}). Also we use Eq.~(\ref{gg}) to get
\be
g=\frac{1}{d^2}
\ee
This result and Eq.~(\ref{thetatind}) allow writting
\be
\label{thetagenind}
\theta=\frac{1}{2d^2}\,\tau
\ee
Eqs.~(\ref{Rgenind}) and (\ref{thetagenind}) give another
solution for the {\it d}-brane in $d>1$ spatial dimensions. We remark that
solutions to the above second-order Eq.~(\ref{Rgphi}) can also be obtained
via the following first-order equation
\be
\label{Rfirst}
\frac{dR}{d\phi}=\frac{1}{d}\,R^{1-d}
\ee
We notice that the above solution (\ref{Rgenind}) solves the first-order
Eq.~(\ref{Rfirst}). We postpone to a future work further investigations
on this and on other related issues.

\section{Comments and conclusions}
\label{sec:comments}

In this paper we have found solutions for the relativistic {\it d}-brane
system in $(d+1,1)$ dimensions for $d=1$ and for $d>1$. The {\it d}-brane
system presents spherically symmetric solutions in $d=2, 3$
that are directly related to the spherically symmetric solutions
introduced in \cite{bja98} for the Galileo invariant system in $(2,1)$
and in $(3,1)$ dimensions. These results were generalized
to higher spatial dimensions, and so we obtained
solutions of the relativistic {\it d}-brane system in $(d+1,1)$ dimensions
that reproduce the solutions $(\ref{sthetad})$ and $(\ref{srhod})$
of the Galileo invariant system $(\ref{actiong})$ in $(d,1)$ dimensions,
for $d>1$.

These solutions are very different from the solutions one finds in $(1,1)$
dimensions. In $d=1$ the Galileo invariant system was solved exactly. The
general behavior follows as in \cite{jpo98}, which solved the equations of
motion following the route of linearization of mechanics of fluids \cite{lli}.
In Ref.~{\cite{jpo98}} one also finds
infinity sets of conserved quantities, which are proper to systems
engendering general integrability. The behavior of general integrability
in $d=1$ shows up in the present investigations via the presence of the wave
equation, which is linear and is solved exactly by standard method.
The 1-brane route to solutions to the Galileo invariant system in $(1,1)$
dimensions is very interesting since it allows obtaining the static solutions
and identifying how the BPS solutions of that dynamical system appear in the
1-brane system.

The investigations done in the present paper introduce further issues,
for instance the problem of searching for other solutions. In connection with
Ref.~{\cite{bja98}} one can ask about the possibility of transforming
solutions to new ones, using the field-dependent coordinate transformations
generated by ${\bf G}$ in Eq.~(\ref{b-}). In the hydrodynamical formulation
of quantum mechanics we can ask about the possibility of not only introducing
diffusion \cite{dgo94} but also going beyond the description of pure
states \cite{fan57}. Other interesting investigations \cite{jpo98a} have been
done, generalizing not only the way the fluid mechanical system interacts
but also how the kinematical contribution may modify its dynamical behavior.

We can also reformulate the light-cone description of the {\it d}-brane in
the way that leads to the Born-Infeld equation \cite{bin34}, as presented in
Refs.~{\cite{bho93,hop94}, for instance. As one knows, the Born-Infeld
equation arises after a nonlinear modification of standard electrodynamics,
and presents interesting properties \cite{ann89} and connections to fluid
mechanics and other issues, as the ones investigated recently in
Refs.~{\cite{cma98,gib98}}. Another interesting issue concerns the fact that
{\it d}-branes are extended objects that can be related to matrix theory and
as such may be of interest for instance in the context of holomorphic
configurations recently considered in Ref.~\cite{cta98}}.
\vskip 1cm

\begin{center}
Acknowledgments
\end{center}

The author would like to thank Roman Jackiw, from whom he has learned quite
a few interesting things related to the subject of this work. He also thanks
Jens Hoppe and Washington Taylor for interesting comments and discussions.
He is grateful to the Center for Theoretical Physics, MIT, for hospitality.


\begin{thebibliography}{30}

\bibitem{bja98}D. Bazeia and R. Jackiw, Ann. Phys. (NY), in press,
hep-th/9803165.

\bibitem{jpo98}R. Jackiw and A.P. Polychronakos, Faddeev {\it Festschrift}
Proceedings, hep-th/9809123.

\bibitem{lli}L. Landau and E. Lifschitz, {\it Fluid Mechanics,} 2nd ed.
(Pergamon, Oxford UK, 1987).

\bibitem{mad26}E. Madelung, Z. Phys. {\bf 40}, 322 (1926).

\bibitem{mer}E. Merzbacher, {\it Quantum Mechanics,} 3rd ed.
(Wiley, New York, 1998).

\bibitem{hop82}J. Hoppe, MIT PhD Thesis (unpublished, 1982).

\bibitem{bho93}M. Bordemann and J. Hoppe, Phys. Lett. B {\bf 317}, 315 (1993).

\bibitem{hop94}J. Hoppe, Phys. Lett. B {\bf 329}, 10 (1994).

\bibitem{hop95}J. Hoppe, preprint, hep-th/9503069.
 
\bibitem{jev98}A. Jevicki, Phys. Rev. D {57}, R5955 (1998).

\bibitem{sch96}A.M.J. Schakel, Mod. Phys. Lett. B {\bf 10}, 999 (1996).

\bibitem{oga98}N. Ogawa, preprint, hep-th/9801115.

\bibitem{sch98}A.M.J. Schakel, preprint, cond-mat/9805152.

\bibitem{fja88}L.D. Faddeev and R. Jackiw, Phys. Rev. Lett. {\bf 60},
1692 (1988).

\bibitem{fan57}U. Fano, Rev. Mod. Phys. {\bf 29}, 74 (1957).

\bibitem{wig32}E. Wigner, Phys. Rev. {\bf 40}, 749 (1932).

\bibitem{cza83}P. Carruthers and F. Zachariasen, Rev. Mod. Phys. {\bf 55},
245 (1983).

\bibitem{dgo94}H.D. Doebner and G.A. Goldin, J. Phys. A {\bf 27},
1771 (1994).

\bibitem{sus68}L. Susskind, Phys. Rev. {\bf 165}, 1535 (1968).

\bibitem{dho98}C. Duval and P.A. Horvathy, unpublished.

\bibitem{hpri}J. Hoppe, private communication.

\bibitem{bps}M.K. Prasad and C.M. Sommerfield, Phys. Rev. Lett. {\bf 35},
760 (1975); E.B. Bogomol'nyi, Sov. J. Nucl. Phys. {\bf 24}, 449 (1976).

\bibitem{jpo98a}R. Jackiw and A.P. Polychronakos, preprint, hep-th/9902024.

\bibitem{bin34}M. Born and L. Infeld, Proc. R. Soc. London A {\bf 144},
425 (1934).

\bibitem{ann89}M. Arik, F. Neyzi, Y. Nutku, P.J. Olver, and J.M. Verosky,
J. Math Phys. {\bf 30}, 1338 (1989).

\bibitem{cma98}C.G. Callan and J.M. Maldacena,
Nucl. Phys. B {\bf 513}, 198 (1998).

\bibitem{gib98}G.W. Gibbons, Nucl. Phys. B {\bf 514}, 603 (1998).

\bibitem{cta98}L. Cornalba and W. Taylor, Nucl. Phys. B {\bf536}, 513 (1998).
\end{thebibliography}
\end{document}